\documentclass[11pt]{article}
\usepackage{jcappub}
\usepackage{array, natbib, amsfonts}
\usepackage{amssymb, amsmath}
\usepackage{epsfig}
\usepackage{graphicx,color}

\newcommand{\rmd}{\ensuremath{\mathrm{d}}}
\newcommand{\hMpc}{{$h^{-1}$Mpc}}

\newcommand{\zl}{{z_l}}
\newcommand{\zp}{{z_p}}

\newcommand{\gammat}{{\gamma_{\rm t}}}
\newcommand{\tgammat}{{\tilde{\gamma}_{\rm t}}}
\newcommand{\tgamma}{{\tilde{\gamma}}}

\newcommand{\gammaIA}{{\gamma^{\rm IA}}}
\newcommand{\gammaIAj}{{\gamma^{\rm IA}_j}}
\newcommand{\gammaIAb}{{\bar{\gamma}_{\rm IA}}}

\newcommand{\sigmac}{{\Sigma_{\rm c}}}
\newcommand{\sigmacj}{{\Sigma_{{\rm c},j}}}
\newcommand{\tsigmac}{{\tilde{\Sigma}_{\rm c}}}
\newcommand{\tsigmacj}{{\tilde{\Sigma}_{{\rm c},j}}}

\newcommand{\DS}{{\Delta\Sigma}}

\newcommand{\mnras}{{\rm MNRAS}}
\newcommand{\apj}{{\rm ApJ}}
\newcommand{\apjs}{{\rm ApJS}}
\newcommand{\araa}{{\rm Annu.Rev.Astron.Astrophys.}}
\newcommand{\jcap}{{\rm JCAP}}
\newcommand{\aap}{{\rm Astron.Astrophys.}}
\newcommand{\nat}{{\rm Nature}}
\newcommand{\prd}{{\rm Phys.Rev.D}}
\newcommand{\apjl}{{\rm ApJL}}
\newcommand{\aj}{{\rm AJ}}
\newcommand{\physrep}{{\rm Phys. Rep.}}

\title{Separating intrinsic alignment and galaxy-galaxy lensing}
\author[a]{Jonathan Blazek,}
\author[b,c]{Rachel Mandelbaum,}
\author[a,d,e,f]{Uro\v{s} Seljak}
\author[g]{and Reiko Nakajima}
\affiliation[a]{Department of Physics and Lawrence Berkeley National Laboratory,\newline University of California, Berkeley, CA 94720, USA}
\affiliation[b]{Peyton Hall Observatory, Princeton University, Peyton Hall, Princeton, NJ 08544, USA}
\affiliation[c]{Department of Physics, Carnegie Mellon University, Pittsburgh, PA 15213, USA}
\affiliation[d]{Department of Astronomy, University of California, Berkeley, CA 94720, USA}
\affiliation[e]{Institute of Theoretical Physics, University of Zurich, Winterthurerstrasse 190,\newline CH-8057, Zurich, Switzerland}
\affiliation[f]{Institute for the Early Universe, Ewha University, Seoul 120-750, South Korea}
\affiliation[g]{Argelander Institut f\"{u}r Astronomie, Auf dem H\"{u}gel 71, 53121 Bonn, Germany}
\emailAdd{blazek@berkeley.edu}

\abstract{The coherent physical alignment of galaxies is an important systematic for gravitational lensing studies as well as a probe of the physical mechanisms involved in galaxy formation and evolution. We develop a formalism for treating this intrinsic alignment (IA) in the context of galaxy-galaxy lensing and present an improved method for measuring IA contamination, which can arise when sources physically associated with the lens are placed behind the lens due to photometric redshift scatter. We apply the technique to recent Sloan Digital Sky Survey (SDSS) measurements of Luminous Red Galaxy lenses and typical ($\sim {\rm L}_*$) source galaxies with photometric redshifts selected from the SDSS imaging data. Compared to previous measurements, this method has the advantage of being fully self-consistent in its treatment of the IA and lensing signals, solving for the two simultaneously. We find an IA signal consistent with zero, placing tight constraints on both the magnitude of the IA effect and its potential contamination to the lensing signal. While these constraints depend on source selection and redshift quality, the method can be applied to any measurement that uses photometric redshifts. We obtain a model-independent upper-limit of roughly 10\% IA contamination for projected separations of $r_p \approx 0.1\text{--}10$ \hMpc. With more stringent photo-$z$ cuts and reasonable assumptions about the physics of intrinsic alignments, this upper limit is reduced to $1\text{--}2\%$. These limits are well below the statistical error of the current lensing measurements. Our results suggest that IA will not present intractable challenges to the next generation of galaxy-galaxy lensing experiments, and the methods presented here should continue to aid in our understanding of alignment processes and in the removal of IA from the lensing signal.}

\arxivnumber{}

\begin{document}
\maketitle
\flushbottom

\section{Introduction}

Observations of gravitational lensing, the deflection of light by matter between the source and observer, have become an important and widely-used tool in cosmology and astrophysics (for a review, see \cite{bartelmann01,refregier03,hoekstra08,massey10}). Lensing measurements are equally sensitive to all types of matter, making them a particularly useful probe of dark matter and theories of gravitation \cite{zhang07}. Galaxy-galaxy lensing measures the gravitational influence of massive foreground (``lens'') objects on the measured shapes of background (``source'') galaxies. In the weak lensing regime, such effects must be studied statistically, since the lensing deflections are typically factors of several tens to hundreds smaller than the intrinsic galaxy ellipticities. Correlating the observed source shapes with lens positions measures the galaxy-mass cross-correlation, which can in turn be used to examine the density profile of halos and probe the standard ${\rm\Lambda CDM}$ paradigm as well as theories of modified gravity \cite{reyes10,lombriser11}. Combining such lensing results with clustering measurements of similar lenses may provide an especially robust probe of density fluctuations \cite{baldauf10}.

Coherent, large-scale correlations of the intrinsic shapes and orientations of observed galaxies are a potentially significant source of systematic error in gravitational lensing studies, with the corresponding potential to bias or degrade lensing science results. This intrinsic alignment (IA) has been examined from both observational \cite{mandelbaum06a,hirata07,okumura09a,okumura09b,joachimi11,mandelbaum11} and theoretical \cite{croft00, heavens00, lee00,lee01, catelan01, crittenden01,jing02,hui02,hirata04,lee08,schneider10,blazek11,schneider11} perspectives. For instance, luminous red galaxies have been observed to align preferentially towards overdense regions. Since lensing increases the observed tangential shape of background galaxies, contamination from such radial alignment would reduce the inferred lensing effect. This type of IA contribution is often referred to as a ``GI'' term, referring to the correlation between matter density, which sources gravitational (G) shear, and intrinsic (I) shear (see \cite{hirata04,blazek11}). It is this type of IA effect that can contaminate a galaxy-galaxy lensing measurement, and one can relate the observed distribution of lens galaxies to the underlying matter distribution to equate with the GI term as originally formulated in the context of cosmic shear. Thus, with sufficient modeling of IA and the effects of redshift quality, a measurement of IA in galaxy-galaxy lensing can probe the IA contribution to cosmic shear.

In the context of galaxy-galaxy lensing, the use of accurate spectroscopic redshifts would ensure that lens and source galaxies are separated by a large distance along the line-of-sight and thus that any intrinsic correlations are negligible. Although such correlations may exist between the shapes of the background source galaxies, they yield no total signal when measurements around many lenses are ``stacked.'' However, obtaining large numbers of spectroscopic redshifts is highly resource intensive, making it difficult to form large samples and achieve small statistical uncertainty. An alternative approach is to use photometric redshifts (photo-$z$'s) from multi-band imaging. Although less accurate than spectroscopic redshifts, photo-$z$'s make much larger galaxy catalogs readily available, greatly improving the potential signal-to-noise of lensing measurements. The degradation of distance information due to the use of photo-$z$'s requires additional calibration during lensing analysis \cite{mandelbaum08,nakajima11} and may introduce additional systematic uncertainties. As many of the next generation surveys will be photometric (e.g., KIDS\footnote{KIlo Degree Survey, http://kids.strw.leidenuniv.nl/}, DES\footnote{Dark Energy Survey, https://www.darkenergysurvey.org}, HSC\footnote{Hyper Suprime-Cam, http://oir.asiaa.sinica.edu.tw/hsc.php}, and LSST\footnote{Large Synoptic Survey Telescope, http://www.lsst.org}), a detailed quantitative understanding of the impact of using photo-$z$'s will be crucial for the future of precision weak lensing.

When using photo-$z$'s in a galaxy-galaxy lensing measurement, IA can contribute a non-negligible signal. Large redshift uncertainties can result in galaxies that are physically associated with the lens being mistakenly assigned a large line-of-sight separation. Thus, if there exists a significant correlation between galaxy shape and the density field, as traced by the location of lens galaxies, IA contamination can enter the lensing signal.

Despite previous work, the physical mechanisms underlying IA, as well as the magnitude of IA in relevant lensing samples, remain poorly understood. In this paper, we develop and apply a technique to measure this IA contamination. Previous studies have used low-redshift spectroscopic observations to directly measure the IA shear - density correlation (e.g. \cite{bernstein02b,sheldon04,hirata07}). In \cite{joachimi11}, a similar measurement is made using photometric redshifts of Luminous Red Galaxies (LRGs) at intermediate redshift ($0.4<z<0.7$), requiring the removal of the lensing effects that contribute due to redshift uncertainty. In \cite{hirata04b}, IA is constrained in the context of galaxy-galaxy lensing under the assumption that no IA contamination exists in a photo-$z$ selected group of background source galaxies. We propose a new method to simultaneously measure the IA and lensing signals from photometric lensing measurements with minimal assumptions.  We exploit the fact that the IA signal is sourced by contamination from galaxies physically associated with the lens. Splitting the source catalog by separation in photo-$z$ from the lens allows us to compare samples with different levels of contamination and thus solve simultaneously for the IA signal and lensing shear. This method is completely self-contained and model-independent, using only the lensing measurements themselves to constrain IA. Limited model-dependent assumptions can also be applied to improve constraints (see \cite{bernstein09,joachimi10,zhang10b,troxel12} for techniques to remove IA using flexible models).

We apply this method to LRG lenses and typical photometric galaxy lenses in the Sloan Digital Sky Survey (SDSS). The main distinctions between this study and \cite{joachimi11} are the lower redshift range for this work, and the fact that \cite{joachimi11} directly measured the intrinsic alignments of LRGs whereas we use the galaxy-galaxy lensing signal to measure the intrinsic alignments of typical source galaxies that dominate lensing measurements. Using this technique with upcoming large imaging surveys will allow us to probe IA across a range of scales and as a function of both lens and source properties. A potentially valuable application of this method is in the galaxy-galaxy lensing mass determination of galaxy clusters. The use of clusters to constrain cosmological parameters relies on understanding and removal of effects which could bias the inferred masses. Better characterization of alignment in different galaxy populations will improve our understanding of systematic errors for weak lensing studies in general, and will provide a probe of the complex astrophysical processes involved in galaxy formation and evolution.

The paper is organized as follows. Section~\ref{sec:observation} describes the observations involved in galaxy-galaxy lensing and the data utilized in this study. Section~\ref{sec:theory} contains a brief overview of the formalism of galaxy-galaxy weak lensing, including IA. Section \ref{sec:method} discusses our method of extracting IA from the lensing signal, and section \ref{sec:results} shows our results. We conclude with a discussion of the implications of our work in section \ref{sec:disc}. An appendix includes a more technical discussion of issues relevant to our measurement method. Calculating the lensing signal and calibration factors requires a cosmological model, although this assumption has a minimal impact on the IA results. Where relevant, we adopt a flat ${\rm\Lambda CDM}$ universe with $\Omega_m = 0.25$ and present results in units with $h=1$.

\section{Data}
\label{sec:observation}

For measurement of galaxy-galaxy lensing and of intrinsic alignments, we use data from the Sloan Digital Sky Survey. A requirement for this study is to have reasonably accurate redshift information for the galaxies that we use as lenses (tracing the density field) and those that we use to measure shapes (tracing either the intrinsic alignments or the lensing shears, depending on the line-of-sight position with respect to the lens). In addition, we must measure galaxy shapes at sufficient resolution. In this section, we describe the data used for the lens and shape samples.

The SDSS \cite{york00} imaged roughly $\pi$ steradians of the sky and spectroscopically followed-up approximately one million of the detected galaxies \cite{eisenstein01, richards02, strauss02}. The imaging was carried out by drift-scanning the sky in photometric conditions \cite{hogg01, ivezi04}, in five bands ($ugriz$) \cite{fukugita96, smith02} using a specially-designed wide-field camera \cite{gunn98}. These imaging data were used to create the source catalog used in this paper. We also use SDSS spectroscopy for the lens galaxies.  All of the data were processed by completely automated pipelines that detect and measure photometric properties of objects, and astrometrically calibrate the data \cite{lupton01, pier03, tucker06}. The SDSS I/II imaging surveys were completed with a seventh data release \cite[DR7,][]{abazajian09}, though this work also relies on an improved data reduction pipeline that was part of the eighth data release, from SDSS III \cite{aihara11}; and an improved photometric calibration \cite[`ubercalibration',][]{padmanabhan08}.

\subsection{Lens sample}

The lens sample considered here consists of LRG lenses from DR7 as selected by \cite{kazin10}. To match cuts in the shape sample on imaging and PSF quality, positions with respect to bright stars, and extinction, we remove 8\% of the area, yielding a final area of 7,131 deg$^2$. The lens sample consists of 62,081 galaxies with spectroscopic redshifts between $0.16\le z <0.36$ and $g$-band absolute magnitudes $M_g$ in the range $[-23.2,-21.2]$, with a roughly constant comoving density of $10^{-4} (h^{-1}\text{Mpc})^{-3}$. For all calculations, we include a weight for each lens designed to mitigate the effects of fiber collisions, completeness, and large-scale structure fluctuations \cite{kazin10}.

\subsection{Shape sample}

The sample of galaxies with shape measurements, used in this paper to study both the gravitational shear and the intrinsic alignment field, is described in \cite{reyes11}.  This catalog is based on the SDSS DR8 photometric data, processed using re-Gaussianization \cite{hirata03} to correct for the effect of the point-spread function (PSF). It contains $\sim 30$ million galaxies ($1.2$ arcmin$^{-2}$) with $r$-band apparent magnitude $r<21.8$. There are also cuts on the image quality, quality of the PSF estimation, galaxy size compared to the PSF, and Galactic extinction. The photometric redshifts assigned to each galaxy based on the five-band photometry are from the Zurich Extragalactic Bayesian Redshift Analyzer \citep[ZEBRA,][]{feldmann06}. The photo-$z$ uncertainty is $\sigma_z/(1+z)\sim 0.11$, due primarily to the low $S/N$ for the majority of the photometric galaxies. The impact of this uncertainty on lensing measurements with this catalog is characterized by \cite{nakajima11}. For this analysis, we use the entire source sample, as well as ``red'' and ``blue'' color sub-samples, where the split is done using the ZEBRA template type (see appendix~\ref{sec:equatingIA} for more details). For comparison with other IA results, note that the typical luminosity of these sources in the redshift range of interest is $\approx {\rm L}_*$, making them comparable to the SDSS L4 sample. Where relevant, we refer to the SDSS luminosity classification scheme described in \cite{tegmark04}, based on $r$-band absolute magnitude. The labels \{L1, L2, L3, L4\} correspond to $M_r$ in the ranges \{[-17,-16]; [-18,-17]; [-19,-18]; [-20,-19]\}.

\section{Lensing formalism}
\label{sec:theory}
As light travels from a distant galaxy, the presence of intervening matter alters its path, acting as a ``gravitational lens'' and altering the observed shape of the galaxy. This gravitational shear is determined by the projected mass density along the line-of-sight. In this section we summarize the relevant aspects of lensing formalism and develop a consistent treatment of intrinsic alignment. For a more detailed treatment of lensing, see, e.g., \cite{bartelmann01} or \cite{mandelbaum05,reyes11} for systematic issues.

\subsection{Galaxy-galaxy lensing}
Observed galaxy shapes are a combination of the intrinsic shape and the gravitational shear. The observed ellipticity of an object, $e$, is related to observed shear by the responsivity factor $\mathcal{R}$, which measures the average response of an ensemble of source shapes to a given shear: $\langle\gamma\rangle=\langle e\rangle/2\mathcal{R}$ \cite{bernstein02}. In the weak lensing limit, one can write the shear as a sum of gravitational (G) and intrinsic (I) contributions: $\gamma = \gamma^{\rm G}+\gamma^{\rm I}$. For an individual object, the intrinsic shape dominates, although if the intrinsic components of galaxy shapes are not correlated with each other, they will average to zero and simply contribute a ``shape noise'' to the measurement. Thus, it is often assumed that the observed shear components provide an unbiased estimator of the true gravitational shear: over a large number of sources $\langle \tilde{\gamma} \rangle = \gamma^{\rm G}$, where a tilde denotes an observed quantity. In the case of galaxy-galaxy lensing, the lensing of background (``source'') galaxies is dominated by a single massive foreground (``lens'') object, which imparts a tangential shear to the observed shapes, yielding a correlation with the lens position. We define the lens surface density contrast, $\DS$, as the difference between the average (projected) surface mass density within radius $r_p$ and the surface mass density at $r_p$: $\DS(r_p)= \bar{\Sigma}(< r_p)-\Sigma(r_p)$. The tangential gravitational shear, $\gamma_{\rm t}^{\rm G}$, is related to the surface density contrast: 
\begin{align}
\label{eq:basiclensing}
\DS=\sigmac \gamma_{\rm t}^{\rm G},
\end{align}
where $\sigmac$ is the critical surface density, expressed in comoving coordinates as
\begin{align}
\sigmac=\frac{c^2}{4\pi G}\frac{D_s}{\left(1+z_l\right)^2 D_l D_{ls}}.
\end{align}
The angular diameter distances $D_l$, $D_s$, and $D_{ls}$ are between the observer and lens, observer and source, and lens and source, respectively, while $z_l$ is the lens redshift.

If the intrinsic shapes of nearby galaxies are correlated with each other, an intrinsic alignment shear contribution, which we denote $\gamma^{\rm IA}$, will be present:
\begin{align}
\label{eq:IA1}
\langle \tgammat \rangle = \gamma_{\rm t}^{\rm G}+\gamma_{\rm t}^{\rm IA}
\end{align}
For notational simplicity, in the remainder of this work we drop the ``t'' subscript - all shears are assumed to refer to the tangential component. A galaxy-galaxy lensing measurement consists of averaging over many lens-source pairs to estimate the average lens density contrast:
\begin{align}
\label{eq:observedDS}
\widetilde{\DS}(r_p)=\frac{B(r_p) \sum\limits_j^{\rm lens} \tilde{w}_j \tsigmacj \tgamma_{j}}{\sum\limits_j^{\rm lens} \tilde{w}_j} = \frac{B(r_p) \sum\limits_j^{\rm lens} \tilde{w}_j \tsigmacj \left(\DS(r_p) \sigmacj^{-1} + \gammaIAj\right)}{\sum\limits_j^{\rm lens} \tilde{w}_j},
\end{align}
where $\tilde{w}_j$ is the weight given to each lens-source pair, and $B(r_p)$ is the boost factor, discussed below, which accounts for sources that are physically associated with the lens. In the summation expressions, ``lens'' denotes a sum over all real lens-source pairs with projected separation $r_p$ in the desired bin, while ``rand'' denotes a sum over random lens-source pairs (where the random lenses are distributed with the same angular and radial selection function as the real lenses). The optimal weight \cite{mandelbaum08} for each pair is a combination of the geometric factor $\sigmacj$, shape noise from the variance in the source ensemble ($e_{\rm rms}$), and the individual object measurement noise ($\sigma_{e,j}$):
\begin{align}
\label{eq:weight}
\tilde{w}_j=\frac{1}{\tsigmacj^2 \left( e_{\rm rms}^2 + \sigma_{e,j}^2\right)}.
\end{align}

\subsection{Accounting for physically associated galaxies}
\begin{figure}[h!]
\begin{center}
\includegraphics[width=\textwidth]{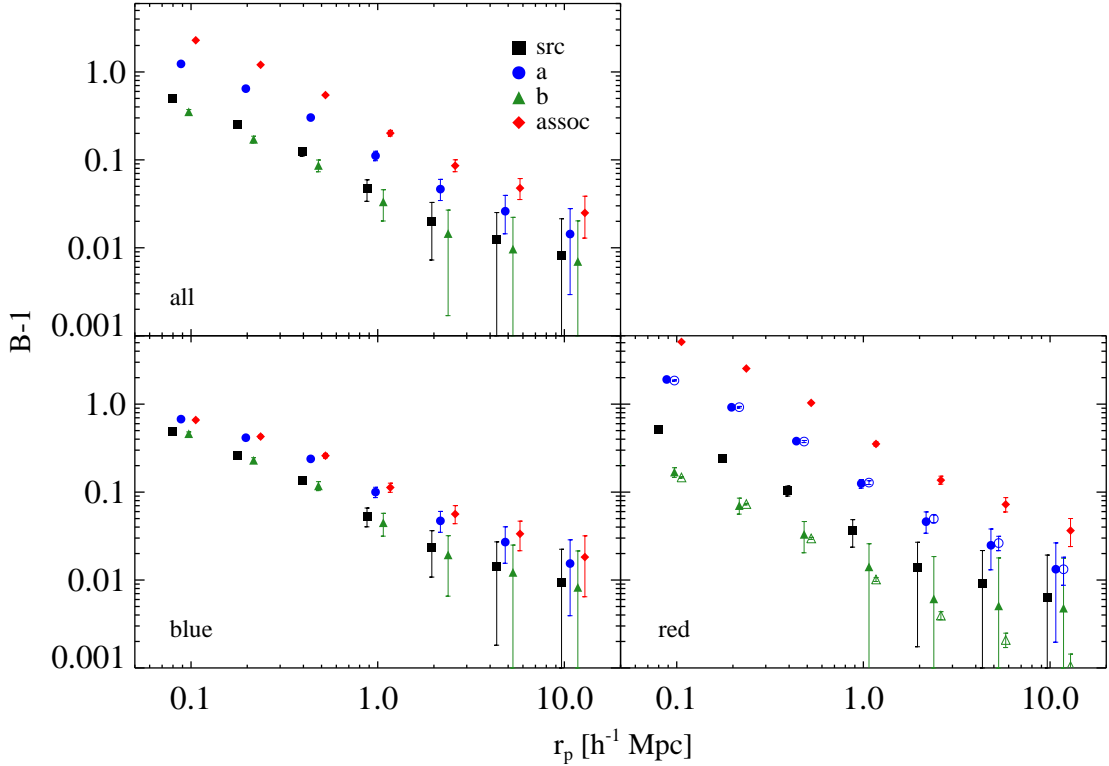}
\end{center}
\caption{Boost factors, which measure the number of ``excess'' lens source pairs due to clustering and photo-$z$ uncertainty, are shown for different source colors and source subsamples (described in sec.~\ref{sec:method}). For red sources, the ``extended boosts'' (described in sec.~\ref{sec:method_2}) are shown as open symbols. As expected, boosts are larger for red galaxies (which cluster more strongly) and for samples defined with photo-$z$'s closer to the lens redshift. In all figures, error bars indicate 68\% confidence regions and small horizontal offsets are added for clarity.}
\label{fig:boosts}
\end{figure}

When looking along the line-of-sight at or near a lens galaxy, the measured number density of sources, as a function of redshift, is given by the sum of two components: a smooth background $\rmd n/\rmd z$, determined by observational parameters and selection cuts; and a sharp peak located at $\zl$ due to galaxies that are physically associated with the lens and thus strongly clustered with it. The boost factor, $B$, measures the contribution of this excess peak in the photometrically defined source sample. Since these excess galaxies are not lensed, they will dilute the measurement unless accounted for explicitly, as done in eq.~\ref{eq:observedDS}. The boost is observationally determined by comparing the weighted number of lens-source pairs for a given source sample with the number of random-source pairs for the same sample:
\begin{align}
\label{eq:boost}
B(r_p)=\frac{\sum\limits_{j}^{\rm lens} \tilde{w}_j}{\sum\limits_j^{\rm rand} \tilde{w}_j}.
\end{align}
With accurate redshift information, it would be easy to exclude physically associated sources, and the boost would approach 1. The measured boosts for the samples used in this study are shown in figure~\ref{fig:boosts}. The quantity $(B-1)/B$ gives the fraction of galaxies in the source sample that are physically associated with the lens. Note that we are assuming that the number of physically associated galaxies is the same as the number of ``excess'' galaxies due to clustering. This assumption need not hold, a subtlety we will explore below.

\subsection{Correcting for photometric redshift bias}

If IA contributes no signal, $\gammaIAj=0$, and the expression in eq.~\ref{eq:observedDS} can be rearranged:
\begin{align}
\label{eq:bzdef}
\frac{\widetilde{\DS}_{\rm IA=0}}{\DS}=\frac{B(r_p)\sum\limits_{j}^{\rm lens} \tilde{w}_j \tsigmacj \sigmacj^{-1}}{\sum\limits_{j}^{\rm lens} \tilde{w}_j}=\frac{\sum\limits_{j}^{\rm rand} \tilde{w}_j \tsigmacj \sigmacj^{-1}}{\sum\limits_{j}^{\rm rand} \tilde{w}_j} \equiv b_z+1,
\end{align}
where we have defined $b_z$, the bias in the lensing signal due to the uncertainty in photometric redshift, and used the fact that excess galaxies are at the lens redshift and thus have $\sigmacj^{-1}\approx\nolinebreak 0$ such that
\begin{align}
\sum\limits_{j}^{\rm lens} \tilde{w}_j \tsigmacj \sigmacj^{-1}=\sum\limits_{j}^{\rm rand} \tilde{w}_j \tsigmacj \sigmacj^{-1}.
\end{align}
Because the lensing efficiency, expressed by $\sigmac$, is a nonlinear function of distance, even an unbiased scatter in redshift can result in a biased lensing measurement, and correction for this photometric bias must be applied when photometric redshifts are used. The value of $b_z$ is a function of the lens redshift, but for the purposes of this study, we only need the average value $\langle b_z\rangle$, determined by integrating over the lens redshift distribution (see, e.g., eq. 22 of \cite{nakajima11}). Combined with the definition of the boost, we can express $b_z$ as a sum over sources around random points rather than lenses, allowing the use of calibration sets from fixed areas of the sky unassociated with the lenses. To calculate $\langle b_z\rangle$, we follow the methods of \cite{nakajima11}, utilizing the same calibration sets. These calibration sets consist of galaxies with both spectroscopic and photometric redshifts, allowing the determination of both $\tsigmacj$ and $\sigmacj$. Object selection in these calibration samples matches that for the source samples considered here.

With these definitions, we can rearrange eq.~\ref{eq:observedDS} to write the true $\DS$ in terms of the observed $\widetilde{\DS}$ and the IA contamination (no longer assumed to be zero):
\begin{align}
\label{eq:DS_total}
\DS(r_p)=c_z\left( \widetilde{\DS}(r_p) - \frac{B(r_p) \sum\limits_{j}^{\rm lens}\tilde{w}_j \gammaIAj\tsigmacj}{\sum\limits_{j}^{\rm lens}\tilde{w}_j }\right),
\end{align}
where $c_z \equiv \left(\langle b_z \rangle+1\right)^{-1}$. In the limiting case of no IA, physically associated galaxies contribute nothing to the measured signal, and the boost acts to cancel this ``dilution.''

\section{Methodology for measuring intrinsic alignment}
\label{sec:method}
\subsection{Solving for IA}
Galaxies that are physically associated with the lens but mistakenly placed behind the lens due to scatter in photometric redshift can lead to IA contamination. At greater line-of-sight separation from the lens, the number of associated galaxies should decrease, assuming a well-behaved photo-$z$ distribution. We exploit this fact to separate IA from the lensing signal. Our IA method is valid in the presence of catastrophic photo-$z$ failures, however the rate and distribution of these failures must be understood using a representative redshift calibration sample. In principle, catastrophic failures could induce a non-physical correlation between IA and observed line-of-sight separation which will bias results if not taken into account. For these SDSS measurements, the failure rate is both low and well-understood and can be safely ignored.

With a minor simplification to the second term in brackets in eq.~\ref{eq:DS_total}, we can explicitly solve for the IA contamination. As discussed above, lens-source pairs can be divided statistically into ``random'' and ``excess'' pairs:
\begin{align}
\label{eq:IAsum}
\sum\limits_{j}^{\rm lens}\tilde{w}_j \gammaIAj \tsigmacj = \sum\limits_{j}^{\rm excess}\tilde{w}_j \gammaIAj \tsigmacj +\sum\limits_{j}^{\rm rand}\tilde{w}_j \gammaIAj \tsigmacj .
\end{align}
We take the sum over ``random'' pairs to be zero, since non-associated pairs should not be aligned with the lens, and it is assumed that the number of excess pairs is a good approximation for the number of physically associated ones. On small scales, where clustering is strong, this assumption is a good one. On larger scales, this assumption may break down, a possibility we address below. We further simplify the expression by replacing the value $\gammaIAj$ for each lens-source pair in the sum with the effective average value for excess pair, $\gammaIAb(r_p)$:
\begin{align}
\label{eq:IAsum2}
\sum\limits_{j}^{\rm lens}\tilde{w}_j \gammaIAj \tsigmacj = \gammaIAb(r_p) \sum\limits_{j}^{\rm excess}\tilde{w}_j \tsigmacj .
\end{align}
Equation \ref{eq:DS_total} can then be rewritten:
\begin{align}
\label{eq:DS_simplified}
\DS(r_p)=c_z \left[ \widetilde{\DS}(r_p) - \left(B(r_p)-1\right)\gammaIAb(r_p) \langle \tsigmac \rangle_{\rm ex}(r_p) \right],
\end{align}
where $\langle \tsigmac \rangle_{\rm ex}(r_p)$ can be directly measured:
\begin{align}
\langle \tsigmac \rangle_{\rm ex}(r_p) = \frac{\sum\limits_{j}^{\rm excess} \tilde{w}_j  \tsigmacj}{\sum\limits_{j}^{\rm excess} \tilde{w}_j} = \frac{\sum\limits_{j}^{\rm all} \tilde{w}_j  \tsigmacj - \sum\limits_{j}^{\rm rand} \tilde{w}_j  \tsigmacj}{\sum\limits_{j}^{\rm all} \tilde{w}_j - \sum\limits_{j}^{\rm rand} \tilde{w}_j}.
\end{align}
This simplification of the IA sum is reasonable because the large photometric uncertainty effectively removes any correlation between observed $\tsigmac$ and $\gammaIA$ for each physically associated object. In other words, the intrinsic alignment shear should depend on the line-of-sight separation between lens and source, but our photo-$z$ errors are far larger than the scales relevant for IA variation.

We consider two groups of lens-source pairs, denoted {\bf a} and {\bf b}, defined in terms of the separation between lens (located at $\zl$) and source (with photometric redshift $\zp$). We take sample {\bf a} to consist of all pairs with $\zl < \zp < \zl +\Delta z$ while sample {\bf b} has $\zp > \zl+ \Delta z$, where $\Delta z=0.17$. Other splitting schemes are feasible, and this technique can be generalized to involve more than two source subsamples, although given the limitations in the statistical power of current measurements, using two subsamples is preferred. The potential advantage of using additional subsamples is discussed in section~\ref{sec:disc}. We assume here that the lens redshift, $\zl$, is determined precisely with spectroscopic measurements, but it is straightforward to include uncertainty in $\zl$. We also refer to the {\bf src} and {\bf assoc} samples, defined by $\zp>\zl$ (i.e. the combined {\bf a} and {\bf b} samples) and $|\zp - \zl|<\sigma_z$, respectively, where $\sigma_z$ is the uncertainty in each galaxy photo-$z$. Samples {\bf a} and {\bf b} are probing the average surface density profile of the same set of lenses, and thus measurements of both come from the same underlying $\DS$.  However, because these samples have different source redshift distributions with respect to the lens positions, they will have different levels of IA contamination: sample {\bf b} consists of lens-source pairs with larger $\zp-\zl$ and should thus have a lower fraction of physically associated galaxies. We further assume that the physically associated galaxies in both samples have the same average value of $\gammaIAb$ and that the two have different levels of IA contamination only due to the different numbers of physically associated galaxies they contain. We discuss this assumption in appendix~\ref{sec:equatingIA}. However, as shown in section~\ref{sec:mimeasure}, potential violations of it produce a sub-dominant bias given current levels of measurement uncertainty, and splitting the source sample by color should largely remove this bias. There are now only two unknown quantities in the lensing measurement: the true $\DS$ and the level of IA contamination per associated object, $\gammaIAb$. With two sets of lens-source pairs, we are able to solve for both $\gammaIAb$ and the true $\DS$ with the IA contamination removed:
\begin{align}
\label{eq:IAsolution}
\gammaIAb(r_p) = \frac{c_{z}^{(a)} \widetilde{\DS}_a-c_{z}^{(b)} \widetilde{\DS}_b}{c_{z}^{(a)} (B_a-1) \langle \tsigmac \rangle_{\rm ex}^{(a)} - c_{z}^{(b)} (B_b-1) \langle \tsigmac \rangle_{\rm ex}^{(b)}},
\end{align}
\begin{align}
\DS(r_p)=\frac{c_{z}^{(a)}c_{z}^{(b)}\left(\widetilde{\DS}_a(B_b-1)\langle \tsigmac \rangle_{\rm ex}^{(b)}-\widetilde{\DS}_b(B_a-1)\langle \tsigmac \rangle_{\rm ex}^{(a)}\right)}{c_{z}^{(b)}(B_b-1)\langle \tsigmac \rangle_{\rm ex}^{(b)}-c_{z}^{(a)}(B_a-1)\langle \tsigmac \rangle_{\rm ex}^{(a)}},
\end{align}
where we have suppressed explicit dependence on $r_p$.

From eq.~\ref{eq:DS_simplified} and the measured $\gammaIAb$, we can calculate the IA contribution to $\DS$, which we denote $\DS_{\rm IA}$, and thus a fractional contamination to the observed lensing signal for a given source sample {\bf s}.
\begin{align}
\label{eq:frac_contam}
\frac{\DS_{\rm IA}}{\widetilde{\DS}}(r_p) = \frac{\gammaIAb(r_p) \langle \tsigmac \rangle_{\rm ex}^{(s)}(r_p)(B_s(r_p)-1)}{\widetilde{\DS}_s(r_p)}.
\end{align}

The weighting applied here for lens-source pairs is the standard scheme used for galaxy-galaxy lensing because it corresponds to weighting by the noise in $\DS$. These are not the optimal weights for measuring the IA signal, since they give less weight to close pairs which are more likely to be physically associated. However, we need to solve for both $\gammaIAb$ and $\DS$. One can solve for the optimal weights to measure each of these quantities, assuming the other is known, and then proceed to solve for both iteratively. However, since the uncertainty in our signal is dominated by the $\widetilde{\DS}$ measurement, the ideal weighting scheme should approach the standard lensing weights. To test a weighting scheme that doesn't favor pairs at large line-of-sight separation, we repeated our measurement using pure shape-noise weighting (removing the factor of $\Sigma_{\rm c}^{-2}$) and found a significantly higher uncertainty in $\gammaIAb$. However, as the precision of future measurements improves, it may be beneficial to use weights optimized for IA measurement with an iterative scheme.

Note that the general technique applied here - comparing the lensing signal observed in two of more samples with different lens-source separations - is similar to a shear-ratio test to probe the geometry of the universe \cite{taylor11}. For the source and lens redshift distributions considered here, we find that varying the cosmological parameters across a reasonable range contributes a highly sub-dominant signal: our choice of cosmology has a negligible impact on IA results. However, for future studies with improved precision, the effects of assuming an incorrect cosmology should be considered. Conversely, if not properly treated, IA has the potential to contaminate shear-ratio or similar measurements and bias the resulting constraints on cosmological parameters.

\subsection{Estimating and reducing uncertainties}
\label{sec:method_2}
To estimate the uncertainties in this measurement, we construct 1000 bootstrap realizations of all measured quantities, by random sampling with replacement, from 100 contiguous regions of the survey footprint. The number of bootstrap regions and realizations is chosen to ensure that the errors are a reasonably smooth function of scale. The quantities $\gammaIAb$ and $\DS_{\rm IA}/\widetilde{\DS}$ are then calculated for each realization, allowing us to directly determine confidence intervals without assuming a Gaussian distribution. Unless otherwise stated, 68\% confidence intervals are shown. We also use these bootstrap realizations to determine model-dependent confidence intervals for IA. See section~\ref{sec:results} for further discussion.

On large scales, where $B \rightarrow 1$, the fractional error in $B-1$ can become quite large. For source subsamples closer to the lens, where the boosts are larger, this fractional error is smaller. If the excess galaxies in different samples have the same composition, these boosts come from the same clustering. The ratio of $B-1$ between these samples should then be constant as a function of scale, with the ratio reflecting the photo-$z$ scatter into each sample. As discussed in appendix~\ref{sec:equatingIA}, such homogeneity between subsamples is present for red galaxies, allowing an improvement in the boost uncertainties. We find the ratio between $B-1$ for different samples on smaller scales ($r_p=0.2\text{--}4.8$\hMpc), where the fractional errors are small. We then take the boosts for the background samples ({\bf a} and {\bf b}) to be the rescaled boosts from the {\bf assoc} sample, thus providing smaller fractional errors. However, as shown in fig.~\ref{fig:gtIA_compare}, this provides only a modest improvement to the overall uncertainty in $\gammaIAb$.

Furthermore, there is a systematic uncertainty in the determination of $B$ of roughly 3\% due to large-scale structure fluctuations and low-level variations in the lens density with observing conditions, which are difficult to model and are not faithfully reproduced in the random catalogs (see Mandelbaum et~al.~2012 {\em in prep.} for details related to this lens catalog, and \cite{ho12} for a related example in SDSS). On scales above $\sim 10$ \hMpc,~${B-1}$ is only a few percent and thus cannot be meaningfully measured. On scales between $\sim 1\text{--}10$ \hMpc, this systematic uncertainty dominates the errors, but the boost values are large enough for accurate measurement. However, this systematic uncertainty leads to some bootstrap realizations having a negative number of ``excess'' lens-source pairs, which is non-physical at these relatively small transverse separations. This artifact could impact the measured uncertainty in $\gammaIAb$ and fractional contamination from IA . To test the impact of this uncertainty, we repeat our calculations using the median boost value rather than that for each individual bootstrap realization, effectively removing the systematic uncertainty, and find little effect on the quantities of interest. We thus conclude that the boost systematic does not significantly influence results on these scales.

\subsection{Identifying physically associated galaxies as ``excess'' galaxies}
Ideally, eq.~\ref{eq:IAsum} would be split into sums over physically associated pairs (which can have IA) and non-associated pairs. However, we are not able to directly measure the fraction of physically associated galaxies in a given sample. Instead, the boost factors determine the fraction of ``excess'' pairs. This excess is similarly described by the two-point cross-correlation between lens and source objects, denoted $\xi_{ls}$. At small separations, where $\xi_{ls} \gg 1$, essentially all associated galaxies are also excess galaxies. On larger scales, however, $\xi_{ls} \lesssim 1$: the number of galaxies that would be present near the lens given a random distribution becomes comparable to and eventually larger than the number of excess galaxies present due to clustering. This complication does not reduce the applicability of the method - the physical mechanism of IA does not distinguish between an object that is excess above random and one that is not. As long as the ratio of excess galaxies to total physically associated galaxies is the same between the two samples, we may simply attribute the entire IA signal to the excess object fraction. The photometric redshift scatter should be, on average, the same for all physically associated galaxies, and thus this ratio should be fixed at a given $r_p$. While this approach allows for accurate IA extraction, it also can lead to seemingly counterintuitive results. For instance, even though alignment should be weaker at larger scales, the value of $\gammaIAb$ may increase above a certain scale, as the number of excess galaxies drops more quickly than the total IA signal (see, e.g., \cite{faltenbacher09}). The integral constraint requires that the average clustering across all scales be zero, while no such constraint exists for the shear correlation signal. This effect should be considered whenever IA is examined in the context of clustering.

\subsection{Combining correlations and photometric redshift uncertainty}
For effective comparison with both theoretical models and previous IA measurements, we wish to express the observables of interest in terms of underlying cosmological quantities. The theoretical values of both the boost factors and $\gammaIAb$ can be written as the convolution of the photometric redshift scatter with the lens-source and the lens-IA cross-correlation functions, $\xi_{ls}=\langle \delta_l \delta_s\rangle$ and $\xi_{l+}=\langle \delta_l \gamma_+\rangle$. Here, $\gamma_+=-\gamma_t$, matching the standard convention used for shear correlation functions.

Consider lenses at $z_l$ and sources at true redshift $z_s$, with photometric redshift $z_p$. For lens and source redshift distributions $p_l(z_l)$ and $p_s(z_s)$ we can calculate the boost for an individual lens at $z_l$:
\begin{align}
\label{eq:B-1 theory}
B(r_p,z_l)-1=\frac{\int \rmd z_s\,p_s(z_s)\xi_{ls}(r_p,\Pi(z_s) ;z_l)\tilde{P}(z_s,z_l)}{\int \rmd z_s\,p_s(z_s)\tilde{P}(z_s,z_l)},
\end{align} 
where $\Pi$ is the line-of-sight separation between the lens and source. $\tilde{P}(z_s,z_l)$ is the weighted probability of an object located at $z_s$ being placed in the source sample due to its assigned $z_p$. This quantity reflects the photometric redshift scatter into the source sample and will depend on the survey properties, the sources under consideration, and the definition of the sample:
\begin{align}
\tilde{P}(z_s,z_l)=\int\limits_{z_{\rm min}}^{z_{\rm max}}\rmd z_p P(z_p|z_s)\tilde{w}(z_p,z_l),  
\end{align}
for a source sample with $z_{\rm min} < z_p < z_{\rm max}$, the boundaries of which may be defined relative to the lens redshift. $P(z_p|z_s)$ is the probability that an object with true redshift $z_s$ is photometrically assigned $z_p$. Here we have assumed that $P(z_p|z_s)$ does not change as a function of distance from the lens due to varying average source properties or observational effects. The total boost factor is found by integrating over the lens distribution $p_l(z_l)$:
\begin{align}
\label{eq:B-1 theory2}
B(r_p)-1=\frac{\int \rmd z_l \,p_l(z_l) \int \rmd z_s\,p_s(z_s)\xi_{ls}(r_p,\Pi(z_s) ;z_l)\tilde{P}(z_s,z_l)}{\int \rmd z_l \,p_l(z_l) \int \rmd z_s\,p_s(z_s)\tilde{P}(z_s,z_l)}.
\end{align}
In an analogous fashion, $\gammaIAb$ can be written in terms of the $\xi_{l+}$ and $\xi_{ls}$:
\begin{align}
\label{eq:gammaIAtheory}
\gammaIAb(r_p)=-\frac{\int \rmd z_l \,p_l(z_l) \int \rmd z_s\,p_s(z_s)\xi_{l+}(r_p,\Pi ;z_l)\tilde{P}(z_s,z_l)}{\int \rmd z_l \,p_l(z_l) \int \rmd z_s\,p_s(z_s)\xi_{ls}(r_p,\Pi ;z_l)\tilde{P}(z_s,z_l)} \approx -\frac{w_{l+}}{w_{ls}}(r_p),
\end{align}
where the minus sign accounts for the convention that positive ``+'' shear corresponds to negative tangential shear, and $w_X$ is the projected correlation function corresponding to $\xi_X$. The quantity $(\xi^{\rm IA}_{l+}(\mathbf{r}))/(1+\xi_{ls}(\mathbf{r}))$ gives the IA contribution to $\gammat$ per lens-source pair (both ``excess'' and ``random'') at $\mathbf{r}$. Different IA models provide predictions for $\xi_{l+}$ which can be tested against observation. As discussed in sec.~\ref{sec:nobackIA}, eq.~\ref{eq:gammaIAtheory} also allows for comparison between the results in this work and previous IA studies.

\section{Results}
\label{sec:results}
\begin{figure}[h!]
\begin{center}
\includegraphics[width=\textwidth]{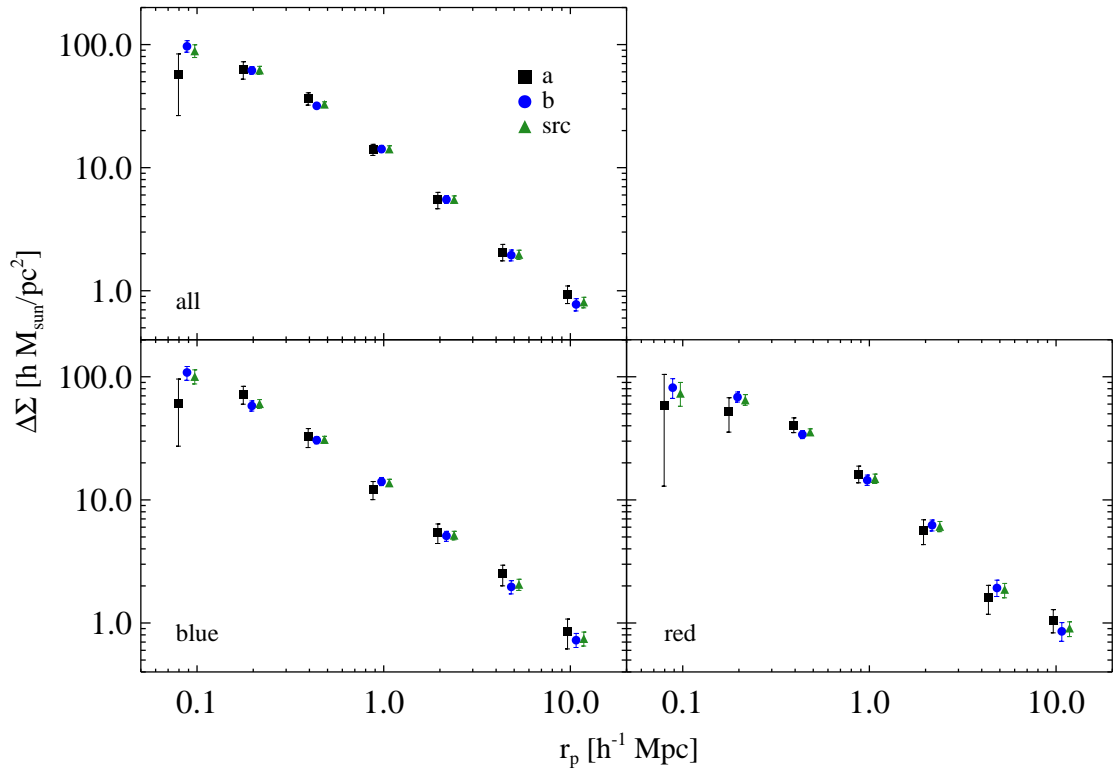}
\end{center}
\caption{The measured $\DS$, including the photo-$z$ bias correction factor $c_z$, is shown for different different source photo-$z$ samples and colors (as labeled). The agreement of the measurements for the different samples indicates that the effects of IA are subdominant at the current level of statistical uncertainty.}
\label{fig:DS}
\end{figure}

We now present the results of the IA measurement technique developed above. We define the {\bf a} and {\bf b} samples using $\Delta z = 0.17$ to yield a roughly equal number of lens-source pairs in the two and minimize measurement uncertainty. See appendix~\ref{sec:uncertainties} for a discussion of this split. Figure~\ref{fig:DS} shows measurements of $\widetilde{\DS}$ for the {\bf a}, {\bf b}, and entire {\bf src} samples.\footnote{Mandelbaum et~al. (2012, {\em in prep.}) will include a more thorough discussion of these results, and cosmological interpretation.} These measurements include the photo-$z$ calibration correction factor $c_z$ for each sample and thus would provide an unbiased estimate of the true $\DS$ in the absence of IA. Any statistically significant deviation between measurements for the {\bf a} and {\bf b} samples would represent either an error in calibration or the presence of IA. It is apparent that the measurements are consistent within the confidence intervals (68\%), so we do not expect a statistically significant detection of IA. Instead, we seek to place tight constraints on the potential IA contribution. We probe projected separations of $r_p= 0.9\text{--}10.1$ \hMpc. On larger scales, the systematic uncertainties in measured boosts, discussed above, greatly degrade the power of our method. On smaller scales, lensing magnification, non-weak shear, and sky systematics in the SDSS software pipeline limit measurement accuracy \cite{mandelbaum06c}.

\subsection{Previous IA measurement methods}
\label{sec:nobackIA}
The work of \cite{hirata04b} attempts to constrain IA by applying similar source photo-$z$ cuts and then assuming that the background sample is effectively free of IA contamination. With this assumption, it is straightforward to estimate the lensing signal in the sample more closely associated with the lens and solve for the IA in that sample. For a background sample, {\bf b}, with no IA:
\begin{align}
\DS(r_p)=c_z^{(b)}\widetilde{\DS}_{b}(r_p).
\end{align}
Thus IA in an associated sample, {\bf a}, can be expressed as
\begin{align}
\gammaIAb(r_p) = \frac{\widetilde{\DS}_{a}(r_p)-\left(\frac{c_z^{b}}{c_z^{a}}\right) \widetilde{\DS}_{b}(r_p)}{(B_{a}(r_p)-1) \langle \sigmac \rangle_{\rm ex}^{a}(r_p)}.
\end{align}

Results using this technique are shown below in figure~\ref{fig:gtIA_compare}. Note, however, that direct comparison with \cite{hirata04b} is difficult. In that work, shear measurements are done directly (rather than using the quantity $\DS$), and the lensing weights and photo-$z$ bias factor are not taken into account. Furthermore, that work uses flux-limited SDSS Main spectroscopic sample galaxies \cite{strauss02} as lenses, rather than LRGs, which will have different IA and clustering properties.

Other studies (e.g. \cite{hirata07,okumura09a,joachimi11}) directly measure IA using $w_{g+}$ of low-redshift SDSS Main spectroscopic sample galaxies as well as higher redshift LRGs (which display a much stronger signal). The use of spectroscopic redshifts greatly reduces the contamination from lensing and dilution from including non-associated pairs. Some relevant results are shown for comparison. It is necessary to convert such measurements to the quantity relevant for galaxy-galaxy lensing using eq.~\ref{eq:gammaIAtheory}. This conversion is non-trivial, and thus any comparison is only approximate. Similarly, \cite{agustsson06} directly measures $\gammaIAb$ for a low-redshift sample of satellites around massive lens galaxies. They find a significant signal below $r_p=0.1$\hMpc, smaller than the scales considered here. On larger scales, their results are consistent with zero IA. The authors of \cite{mandelbaum11} constrain IA for higher-$z$ blue galaxies ($z\sim 0.6$) using shapes from SDSS and spectra from the WiggleZ survey \cite{glazebrook07}. Their results are consistent with the limits we find here.

\subsection{Model-independent measurement}
\label{sec:mimeasure}
\begin{figure}[h!]
\begin{center}
\includegraphics[width=\textwidth]{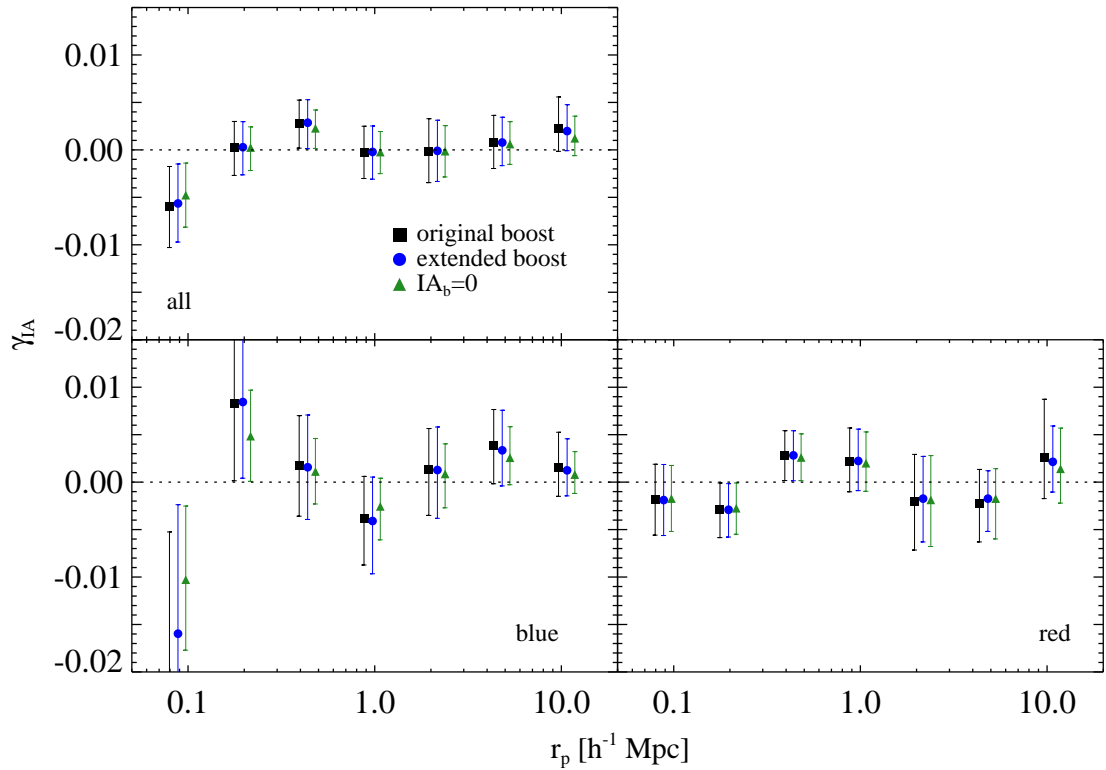}
\end{center}
\caption{The measured values of $\gammaIAb$ using three different techniques are shown (with each panel showing the results for a different source color sample). Black squares indicate that the original boost measurements have been used in eq.~\ref{eq:IAsolution}. Blue circles indicate the use of ``extended boosts.'' Green triangles indicate that the IA contamination in the background sample was assumed to be zero. All three methods yield results consistent within the statistical uncertainty, although assuming that the background sample has no IA slightly biases the magnitude of $\gammaIAb$ to lower values.}
\label{fig:gtIA_compare}
\end{figure}

Figure~\ref{fig:gtIA_compare} shows the results of our IA measurement technique across a range of projected separations. For comparison, results are shown with and without the ``extended boosts'' technique discussed in section~\ref{sec:method_2}. Using extended boosts lowers the statistical uncertainty at large scales, although the difference is not significant. Also shown are the results assuming no IA in the background source sample ({\bf b}). This measurement is more analogous to that made in \cite{hirata04b}, although as noted above, direct comparison with these earlier measurements is challenging. As expected, neglecting the IA contribution in the background sample biases the resulting IA measurement to a lower magnitude: the disparity between $\widetilde{\DS}$ measurements in the two samples is effectively increased when allowing for IA in the background sample. The induced bias is smaller than the level of uncertainty in the measurement for blue, red, and all sources. Because this assumption also reduces the statistical uncertainty in the measurement, the resulting confidence intervals are fully contained within those of our more general and self-consistent measurement. Although ignoring IA in the background sample does not bias the IA measurement in a statistically significant way, it results in overly optimistic confidence regions. In effect, neglecting IA in the background sample has converted a statistical error into a systematic error. For the remainder of this work, we consider results using our fully consistent method. For red sources, where the assumption of a uniform excess source population between the two samples is justified, we use the ``extended boosts,'' while for all and blue sources, we use the boosts as measured.

Comparing measurements with and without accounting for background IA tests the impact of the assumption that $\gammaIAb$ is the same in the two samples. As discussed in appendix~\ref{sec:equatingIA}, the {\bf a} and {\bf b} samples can have somewhat different average galaxy alignment properties. However, even if $\gammaIAb$ differs between them, it cannot do so more strongly than in the case of completely neglecting IA in the background sample.\footnote{In principle, the sign of the IA effect could be different in the two source subsamples, although such a scenario is contrived.} Since the bias induced in this limiting case is minimal, we can be confident in applying our measurement method to all, red, and blue sources.

In all cases, IA contamination is consistent with zero at the scales considered. The uncertainty in IA for red galaxies is smaller than for blue galaxies since the former cluster more strongly and have better photo-$z$ precision, leading to a larger difference in the number of excess galaxies between the {\bf a} and {\bf b} samples. Appendix~\ref{sec:uncertainties} discusses the sources of uncertainty and the effects of splitting sources by color and photo-$z$.

\subsection{Including minimal model dependence}
The results shown in the previous section place constraints on IA as a function of projected separation $r_p$, utilizing only the lensing measurements themselves. Treating the IA measurements at different separations as independent is the maximally conservative approach. Including information about the underlying model of IA, which relates the signal strength on different transverse scales, will yield tighter constraints. Determining an accurate model for IA is an ongoing theoretical challenge, made more complicated by the heterogenous nature of source samples used in lensing studies. For this reason, we aim to make fairly minimal assumptions and show the resulting constraints for two cases: a generalized power-law model and a model motivated from IA measurements of LRGs.

\begin{figure}[t!]
\begin{center}
\includegraphics[width=\textwidth]{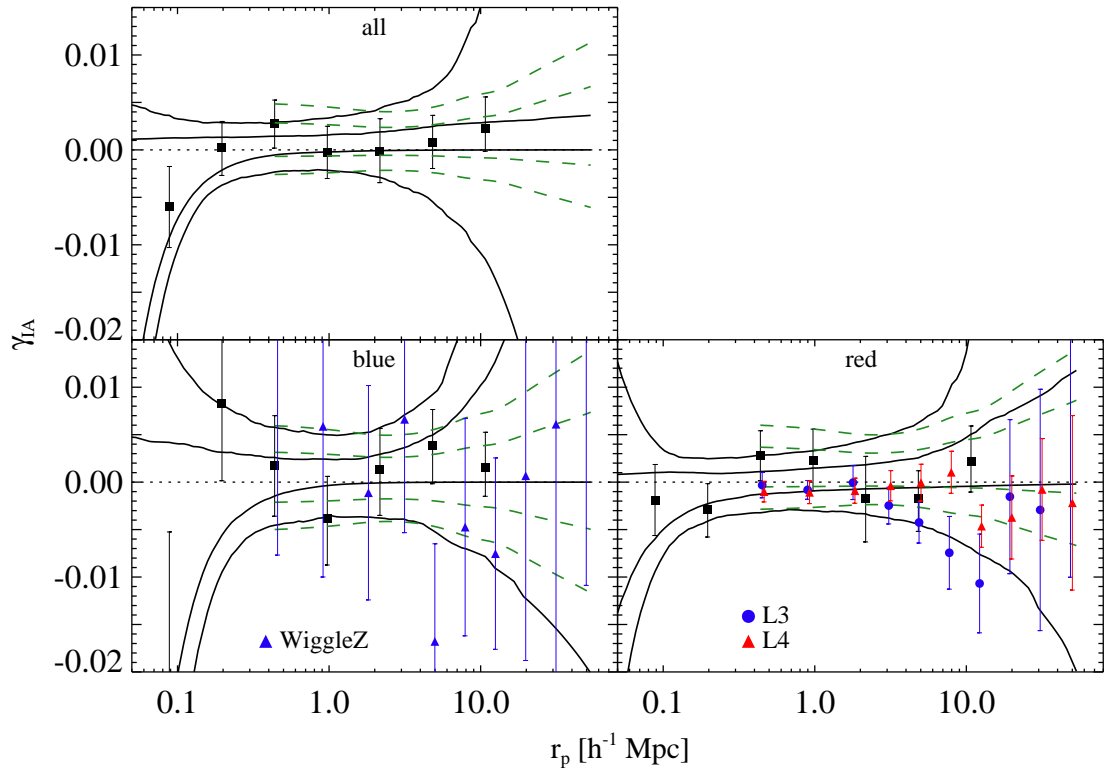}
\end{center}
\caption{Model-dependent confidence intervals for the IA signal for all, blue, and red sources are shown with the model-independent measurements (black data points). Our results are consistent with zero IA signal. Solid black (dashed green) lines denote the power-law (LRG observational) model. Inner lines bound the 68\% confidence region while outer lines correspond to 95\%. Previous observational results for L3 and L4 red galaxies from \cite{hirata07} and for WiggleZ galaxies from \cite{mandelbaum11} are shown for comparison. These previous results have been converted to the $\gammaIAb$ quantity, as discussed in section~\ref{sec:mimeasure}.}
\label{fig:modelCL}
\end{figure}

The 1000 bootstrap realizations are combined to construct a full covariance matrix for the IA measurement:
\begin{align}
\hat{C}_{ij} = \frac{1}{N_{\rm boot}-1}\sum\limits_{k=1}^{N_{\rm boot}}\left[\gammaIAb(r_i)_k\gammaIAb(r_j)_k-\langle\gammaIAb(r_i)\rangle\langle\gammaIAb(r_j)\rangle\right],
\end{align}
where angled brackets indicate an average over the realizations. We then use this covariance matrix to find best-fit parameters for a particular model for each bootstrap realization. At every value of $r_p$, a confidence interval is constructed using the $N_{\rm boot}$ predictions from the model fits. Thus, in the case of a model with multiple parameters or one that does not monotonically change with its parameter, the resulting confidence region envelopes may have a shape different from the model itself. Because it is calculated from bootstrap realizations, the covariance matrix estimated here, $\hat{\boldsymbol{C}}$, is only an estimate of the true covariance matrix and will not result in a $\chi^2$-distribution (see \cite{hirata04b}). However, since we directly construct confidence intervals, we only minimize $\chi^2$ and need not worry about its distribution. After calculating $\hat{\boldsymbol{C}}$, we make the further assumption that off-diagonal terms are zero. As discussed in Mandelbaum et al. (2012, {\em in prep}), shape noise dominates the covariance of the lensing signal on the scales used here, and thus correlations between bins become appreciable only when the same sources are correlated with multiple lenses - i.e., $r_p$ that is roughly twice the typical separation between lenses, or $\approx$ 20\hMpc. Figure~\ref{fig:modelCL} shows the resulting confidence intervals.

We first consider a generalized power-law model of the form $\gammaIAb = A r_p^{\beta}$, where both $A$ and $\beta$ are free parameters fit to each realization. This model is well-motivated in the regime where both $w_{g+}$ and $w_{gg}$ can be approximated by power laws (see, e.g., \cite{hirata07}). As discussed above, the resulting envelope of the confidence region does not follow a power-law. Instead, constraints are tight on the intermediate scales probed by these lensing measurements (where IA signal is consistent with zero), and rapidly increase on large and small scales where there is little IA information.

We also consider a model motivated by the IA observations of LRGs. Following eq.~\ref{eq:gammaIAtheory}, $\gammaIAb$ can be estimated as the ratio $w_{l+}/w_{ls}$. Using the LRG observations of \cite{hirata07}, we calculate this quantity, smoothing with a Savitsky-Golay filter and interpolating to obtain a continuous scale-dependence. We fit the result to our IA measurements with a single parameter for the amplitude, which allows for differences in IA strength and galaxy biasing between the LRGs and the fainter source galaxies used in this work. We use LRG observations because they exhibit a significantly stronger IA signal, providing sufficient $S/N$ (above $r_p=0.9$\hMpc) to determine a well-defined scale dependence. Non-linear effects and environmental dependence of IA mean that LRG measurements on smaller scales are unlikely to accurately reflect the behavior of the relevant sources. We thus use a spline to extend to smaller scales, and note that constraints in this region are contingent on the smooth continuation of IA behavior. We fit these constraints using scales above $r_p=0.44$\hMpc, unlike in the case of the power-law shape, where all measurements are used.

This observational model assumes that the dimmer source galaxies considered here have the same IA scale-dependence as the LRGs, allowing for changes in amplitude due to differences in clustering bias, alignment bias, and object ellipticity. The model is thus of limited use on scales where IA and clustering are qualitatively different (e.g. a different physical mechanism for alignment is dominant). Unlike the power-law model, this observational model provides well-behaved confidence regions on larger scales because LRG observations extend beyond the region where we obtain constraints. As noted above, however, constraints from the observational model on scales below $r_p=0.9$\hMpc\ are less robust.

On larger scales, where clustering is weak, the statistical distinction between ``excess'' and ``associated'' pairs becomes significant, and the number of associated pairs will no longer be well-approximated by $w_{gg}$. In this regime, the validity of both models discussed here is diminished.

\subsection{Contamination to lensing signal}
\begin{figure}[h!]
\begin{center}
\resizebox{\hsize}{!}{
\includegraphics{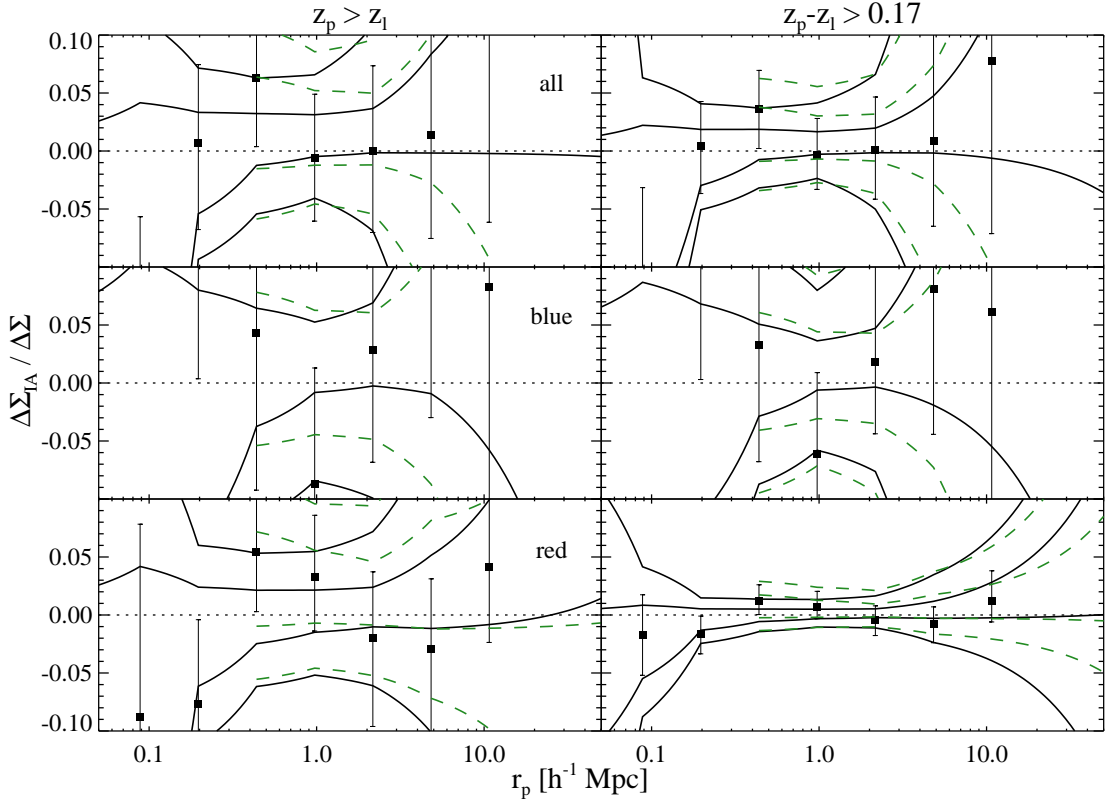}
}
\end{center}
\caption{Fractional contamination to the lensing signal is shown for all, blue, and red galaxies. Data points and lines follow the same convention as figure~\ref{fig:modelCL}. The left column shows the contamination when the entire {\bf src} sample ($z_p>z_l$) is used, while the right column shows contamination when only the background sources at greater separation ($z_p>z_l+\Delta z$) are used. As expected, including scale-dependent assumptions and applying the photo-$z$ cut both provide tighter constraints.}
\label{fig:frac_contam}
\end{figure}

As seen in eq.~\ref{eq:frac_contam}, the level of IA contamination in the lensing signal depends on both the IA signal per excess object, $\gammaIAb$, and the fraction of excess galaxies in the particular sample (directly measured by the boosts). Figure~\ref{fig:frac_contam} shows the fractional contamination resulting from the IA constraints, both with and without assumptions on IA scale-dependence. Results are shown both for all background sources and for a subset of sources with photo-$z$ at least $\Delta z=0.17$ behind the lens. Using this background sample with a photo-$z$ margin removes a large fraction of excess galaxies and thus greatly reduces the potential contamination from IA. Due to the lensing weights used, this cut can be applied without significantly reducing the statistical power of the measurement. With the current level of binning in projected separation, the statistical uncertainty in the $\DS$ measurement is $\approx 5\text{--}15 \%$, depending on the sample and projected separation. For instance, at $r_p=1$\hMpc~the uncertainty in $\DS$ is $\approx 7\%$ for all and blue sources and $\approx10\%$ for red sources. In most cases, the limits on fractional contamination from IA for the scales we measure ($\approx 0.1\text{--}10$\hMpc) are below this uncertainty, significantly so when model dependence is included.

\begin{figure}[h!]
\begin{center}
\resizebox{\hsize}{!}{
\includegraphics{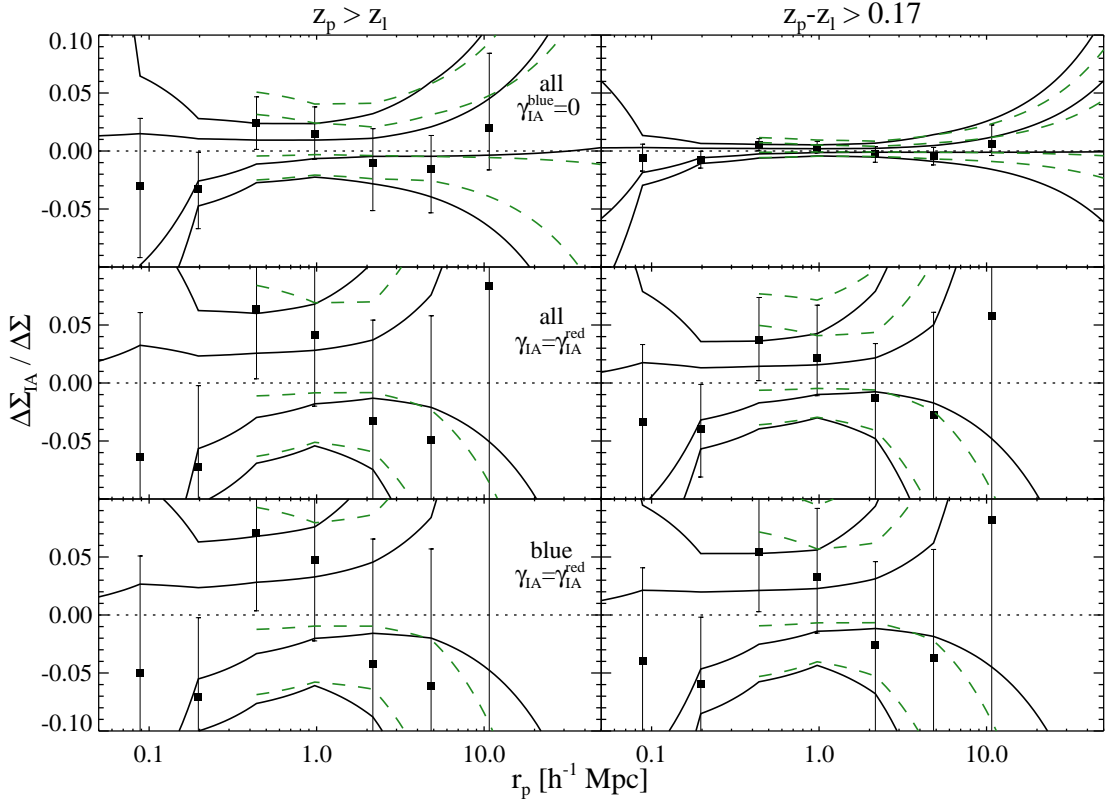}
}
\end{center}
\caption{Fractional contamination to the lensing signal is shown making two different assumptions on $\gammaIAb$. The top row shows the fractional contamination to all sources if $\gammaIAb=0$ for blue galaxies. The middle and bottom rows show the fractional contamination to all and blue sources, respectively, if $\gammaIAb$ for all galaxies is the same as for red galaxies. These reasonable assumptions significantly tighten the constraints on contamination from IA. Columns and line conventions are the same as in figure~\ref{fig:frac_contam}.}
\label{fig:frac_contam_imp}
\end{figure}

The limits on fractional contamination in figure~\ref{fig:frac_contam} for the all and blue samples are conservative. As discussed in section~\ref{sec:mimeasure} and appendix~\ref{sec:uncertainties}, the uncertainty in $\gammaIAb$ is larger for blue than for red galaxies, even though both observational results and theoretical predictions (e.g. \cite{hirata07,faltenbacher09,hirata04}) predict that IA for blue objects is minimal. This consideration suggests that more realistic constraints can be obtained by taking $\gammaIAb$ for blue galaxies to be less than that for red galaxies. Figure~\ref{fig:frac_contam_imp} shows the resulting fractional contamination for two limiting cases: $\gammaIAb=0$ for blue galaxies and $\gammaIAb$ is the same for blue and red galaxies. With the photo-$z$ cut, the more optimistic assumption ($\gammaIAb=0$ for blue galaxies) yields a constraint of $\sim 1\%$ on IA contamination for all source galaxies on the scales we measure. The more pessimistic assumption of $\gammaIAb$ being the same for red and blue galaxies results in IA constraints of a few percent on these scales.

\section{Discussion}
\label{sec:disc}
In this work we have developed a method for measuring the intrinsic alignment of source galaxies in a galaxy-galaxy lensing measurement, as well as the resulting contamination to the lensing signal. Unlike previous IA measurement techniques, the method is fully self-consistent, yielding an unbiased measurement of both IA and $\DS$, and can be used to directly infer contamination to the galaxy-galaxy lensing signal. Applying this method to SDSS LRG lenses and photometric sources, we find a signal consistent with zero IA for all, blue, and red source galaxies at transverse separations of $\approx 0.1\text{--}10$\hMpc. To obtain tighter constraints on IA, we have assumed a functional form for its scale-dependence, using two different physically reasonable models: a general power law and a model motivated by LRG observations. These two models yield similar results on intermediate scales, diverging on large scales where there are no measurements to provide information. We note that while our results are only directly applicable to galaxy-galaxy lensing, the alignment mechanisms being probed are the same as those that can contaminate shear-shear correlations. Given an IA model or parametrization, one could use galaxy-galaxy lensing results to determine constraints on the corresponding GI contamination in shear-shear correlations. The GI term is believed to be the primary IA systematic, since the intrinsic-intrinsic (II) term can be removed with redshift information. Our current work examines a redshift range somewhat below that most relevant for upcoming lensing surveys. Similar analysis with higher redshift lenses will allow us to place constraints on the GI term in future measurements of both galaxy-galaxy lensing and shear-shear correlations.

IA for blue galaxies is less well constrained due to their weaker clustering and poorer photo-$z$ precision. However, previous studies indicate that red galaxies should have a stronger IA signal. Thus, although a conservative approach allows for larger IA in blue galaxies, it is reasonable to take the results for red galaxies as an overall upper limit.

Compared to previous IA measurements, our analysis most closely probes the redshifts and galaxy luminosities relevant to current and future lensing measurements. Our results are broadly consistent with previous studies (e.g. \cite{hirata04b,agustsson06,hirata07,mandelbaum11,joachimi11}), which find insignificant IA signal for similar galaxies at lower redshifts (see figure~\ref{fig:modelCL}). One possible exception is the red L3 sample of \cite{hirata07,joachimi11}, where there is a weak detection of IA at $r_p \approx 10$ \hMpc. However, there are notable differences between our current study and previous ones. We measure IA for typical sources at higher redshift than \cite{hirata04b,hirata07,joachimi11}, and probe alignment in cross-correlation with LRGs rather than in auto-correlation. We also note that recent measurements \cite{hirata07,okumura09a,okumura09b,joachimi11} detect a significant autocorrelation IA signal for an LRG sample similar to our lens sample. This signal is not inconsistent with our findings, which constrain the IA for significantly less luminous galaxies around LRGs. However, this disparity indicates that different types of galaxies display different alignment in the same underlying density field (e.g. LRGs are significantly more aligned). Such a result is not surprising given the likelihood of environmental and nonlinear effects: alignment on these scales depends on more than simply the density field. Comparison with previous IA measurements is difficult since it requires assumptions on how IA and clustering behaviors change on nonlinear scales with object type, redshift, and environment. In addition to effects discussed above, if the primary alignment mechanism acts during periods of formation or accretion, followed by decreasing alignment due to stochastic nonlinear astrophysics, we would also expect an IA signal with significant redshift evolution (see, e.g., \cite{blazek11}). Furthermore, measuring IA at low redshift can be done using spectroscopic galaxies directly without having to account for the lensing signal, as is done in \cite{hirata07}. Any tension between the constraints found here and previous low-$z$ measurements is therefore relevant to IA modeling but not worrisome from a consistency standpoint. We expect future measurements with higher $S/N$ to detect a non-zero IA signal for galaxies typically used as lensing sources, although it will be weaker than that for LRGs.

We also compare our constraints with those found in \cite{mandelbaum11}, who use spectroscopic redshifts to directly constrain the IA of galaxies in the WiggleZ sample. These galaxies have a similar luminosity and roughly comparable redshift distribution to those in this study, although the WiggleZ galaxies peak at larger $z$. The biggest difference between the samples is color: WiggleZ selects emission-line galaxies with UV observations, resulting in a sample that corresponds to the 10--20\% bluest galaxies of the ``blue'' galaxies examined here. Nevertheless, comparison between the results is informative. As seen in figure~\ref{fig:modelCL}, despite the fact that our method is less efficient with blue galaxies than with red galaxies, the constraints we find for this sub-sample are tighter than the limits from \cite{mandelbaum11}. Even allowing for uncertainty in converting between the different types of measurements and different object properties, our photometric approach to IA is competitive with spectroscopic methods, which necessarily use smaller galaxy samples. Given the wealth of upcoming photometric surveys, the disparity in size between photometric and spectroscopic samples with reliable shape measurements will only increase.

We have placed constraints on the potential contamination to the lensing signal from IA: including minimal model assumptions on scale-dependence, $\lesssim 5\%$ contamination is found at the 95\% confidence level on the scales we probe. IA thus remains a subdominant source of uncertainty when compared with the current statistical error on the galaxy-galaxy lensing signal of $\approx$~5--15\%, given the binning in transverse separation. These results apply to the source selection and redshift quality of this particular study - contamination will be worse if less redshift information is used. Constraints on contamination depend not only on source properties such as color, but also on the photo-$z$ selection cut used. As seen in figure~\ref{fig:frac_contam}, constraints are significantly tighter when sources are selected with a photo-$z$ cut to remove a large fraction of galaxies physically associated with the lens. In this work, we apply the cut $z_p > z_l + \Delta z$, for $\Delta z = 0.17$, although optimizing this selection will depend on the specifics of a particular survey and science goals.  Applying such a cut can greatly reduce the possible IA contamination without sacrificing much signal, since distant lens-source pairs dominate the signal due to the weights. Statistical uncertainties in the $\DS$ measurement with and without the photo-$z$ cut agree to within 10\%. Similarly, the relative importance of IA is greater for lens galaxies at higher redshifts: the distribution of background source galaxies will be concentrated closer to the lens positions, increasing the number of physically associated galaxies that receive appreciable weight.

Our findings do not suggest a clear method to cut sources by color to reduce potential IA contamination. Although the constraints found here are tighter for red galaxies, this is because our method performs better for sources that have stronger clustering and better photo-$z$ precision. It is likely that blue galaxies display weaker alignment, and it is unclear to what extent this trend is offset by poorer photo-$z$ precision. Future lensing measurements with lower statistical uncertainty should resolve this issue. However, it is likely beneficial to remove sources with particularly high photo-$z$ uncertainty in order to decrease both potential IA contamination and systematic uncertainty from imperfect photo-$z$ bias correction. In this work, we split the source galaxies into only two sub-samples. With improved statistics, it will be possible to use additional sub-samples in order to break a possible degeneracy between IA strength and correlation between the IA properties and photometric redshift uncertainties of source galaxies. This issue is discussed in appendix~\ref{sec:equatingIA}.

High-precision photometric lensing science will be a primary tool in the future of observational cosmology. As such, upcoming surveys such as KIDS, DES, HSC, and LSST are all designed to obtain photometry and high-resolution imaging for a large number of sources. While the data sets are being built and understood, early focus will likely be on galaxy-galaxy lensing, utilizing the large number of massive foreground galaxies with well-determined redshifts, such as galaxies from current spectroscopic surveys or galaxy clusters. In the near term, galaxy-galaxy lensing analysis is currently ongoing with the BOSS CMASS sample \cite{white11,anderson12} as lenses, and we hope to apply our method as part of the analysis. This sample is at higher redshift ($0.4\lesssim z \lesssim 0.7$), allowing a probe of IA in a new redshift regime relevant for the next generation of lensing surveys and where lens-source overlap may increase the impact of IA. Similarly, we intend to use our method with cluster lenses to study IA of galaxies around cluster centers. Since clusters are even more highly biased than LRGs, we would expect larger numbers of physically associated galaxies in the source sample as well as a higher amplitude for alignment effects. IA contamination for cluster lenses could bias the measured lensing masses and the resulting cosmological results, particularly given that cluster lensing masses are typically determined using lensing signals on even smaller scales than were considered here, where IA effects will be more important.

In the longer term, future studies should obtain galaxy-galaxy lensing signals with uncertainty at the $\sim 1\%$ level. Thus, although IA may become a significant systematic as the statistical errors decrease, applying the method outlined here will yield correspondingly stronger limits on IA (roughly equal to the precision of the lensing measurement) and will allow the removal of this contamination. When IA constraints are improved by assuming a specific radial scaling of the IA signal, they should reach the sub-percent level. Indeed, given the detection of IA in previous spectroscopic studies, these upcoming photometric surveys should achieve a significant detection of IA with the joint lensing and IA measurement method introduced in this paper. These measurements will further reduce systematic uncertainties in lensing studies. Moreover, they will probe the physical mechanisms of intrinsic alignment and thus improve our understanding of galaxy formation and evolution.

\acknowledgments
We thank Matt McQuinn, Teppei Okumura, Andreu Font, Chris Blake, Michael Schneider, and Lucas Lombriser for useful discussions. We also thank an anonymous referee for helpful suggestions. J.B. appreciates the hospitality of the Institute for Theoretical Physics at the University of Zurich, where part of this work was done.

This work is supported by the DOE, the Swiss National Foundation under contract 200021-116696/1, and WCU grant R32-2009-000-10130-0.  

Funding for the SDSS and SDSS-II has been provided by the Alfred P. Sloan Foundation, the Participating Institutions, the National Science Foundation, the U.S. Department of Energy, the National Aeronautics and Space Administration, the Japanese Monbukagakusho, the Max Planck Society, and the Higher Education Funding Council for England. The SDSS Web Site is http://www.sdss.org/.

The SDSS is managed by the Astrophysical Research Consortium for the Participating Institutions. The Participating Institutions are the American Museum of Natural History, Astrophysical Institute Potsdam, University of Basel, University of Cambridge, Case Western Reserve University, University of Chicago, Drexel University, Fermilab, the Institute for Advanced Study, the Japan Participation Group, Johns Hopkins University, the Joint Institute for Nuclear Astrophysics, the Kavli Institute for Particle Astrophysics and Cosmology, the Korean Scientist Group, the Chinese Academy of Sciences (LAMOST), Los Alamos National Laboratory, the Max-Planck-Institute for Astronomy (MPIA), the Max-Planck-Institute for Astrophysics (MPA), New Mexico State University, Ohio State University, University of Pittsburgh, University of Portsmouth, Princeton University, the United States Naval Observatory, and the University of Washington.

\appendix

\section{Equating IA in different samples}
\label{sec:equatingIA}
In principle, solving simultaneously for $\DS$ and $\gammaIAb$ requires that the average value of $\gammaIAb$ in the two source sub-samples, split by photo-$z$ in relation to the lens position, be the same. As discussed above, we find that potential violations of this assumption lead to sub-dominant bias effects given the current level of measurement uncertainty. However, upcoming studies will provide greatly improved precision, and thus a discussion of this issue is warranted.

The above assumption need not hold if some set of galaxy properties, such as color or luminosity, correlates with both the photo-$z$ uncertainty and the level of IA. Source samples defined by different lens-source line-of-sight separation could then have different average IA properties. Since an important step in the determination of an object's photometric redshift is fitting to an expected spectral template based on multi-band photometry, different morphological types can have with different photo-$z$ precision (see \cite{nakajima11}). Both observational and theoretical studies have suggested a divide in IA properties along morphological lines: late-type spirals (typically blue and star-forming) and early-type ellipticals (typically red) are likely subject to differing physical processes that affect alignment, e.g. \cite{hirata04}. In the case of spirals, angular momentum provides the major source of object support and can thus significantly influence orientation via tidal torquing. Elliptical galaxies are pressure-supported through velocity dispersion and are thus expected to align more closely with the surrounding halo and underlying tidal field. Indeed, several observations of IA have found much stronger signal for red galaxies (\cite{hirata07,faltenbacher09}).

To mitigate this issue, one can split the sample by photometric template type. Ideally, multiple splits would be used to isolate individual template types. However, dividing the source sample reduces statistical power, which quickly becomes the limiting factor. Previous IA studies have relied on a simple division between red and blue galaxies, and future lensing measurements are unlikely to consider more complicated template separation. Thus to both maximize the signal and make the results directly relevant, in this work we have applied a single split. The ZEBRA pipeline interpolates between six primary template types, allowing five subdivisions between each. We define all galaxies assigned to the first five interpolated templates, between the elliptical (Ell)  and Sbc primary templates, to be red. All other galaxies are blue (see \cite{nakajima11} for further information on the templates).

\begin{figure}[h!]
\begin{center}
\includegraphics[width=.7\textwidth]{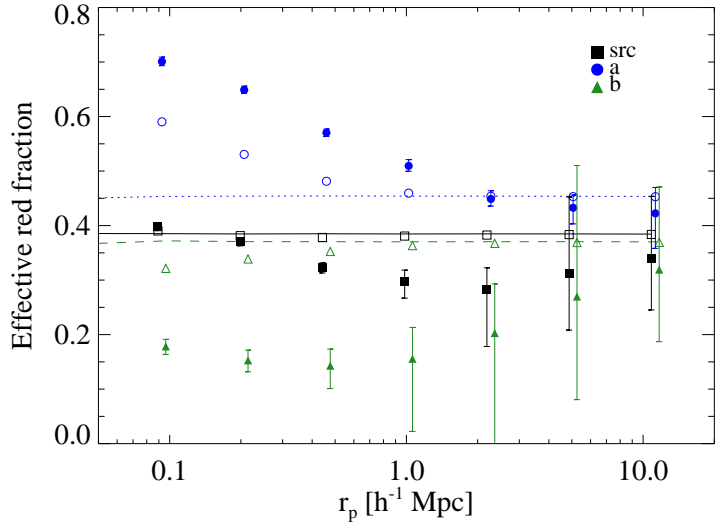}
\end{center}
\caption{The effective fraction of red galaxies is shown for four different source subsamples. The lines (nearly constant) show $f_r$ for {\bf src} (solid), {\bf a} (dotted), and {\bf b} (dashed) samples. Open symbols denote $f_t$, while filled symbols denote $f_e$. Errors in $f_r$ and $f_t$ are negligible and are thus not shown. The composition of the {\bf a} and {\bf b} samples are clearly different.}
\label{fig:eff_frac}
\end{figure}

The sums over weights that determine the boosts can also be used to calculate the effective fraction of red (or blue) galaxies in each source sample as a function of $r_p$, automatically accounting for the lensing weights. Figure~\ref{fig:eff_frac} shows the effective red fraction of galaxies around random lenses ($f_r$), total galaxies around real lenses ($f_t$), and excess galaxies around real lenses ($f_e$). This fraction is defined as $\sum_{\rm red}\tilde{w}_j/\sum_{\rm all}\tilde{w}_j$, where the sums are taken over the set of random, total, or excess lens-source pairs (determined statistically). Because it reflects the underlying distribution and photo-$z$ scatter of sources, with no influence from clustering, $f_r$ should be constant as a function of $r_p$ and is included as a reference and check on systematics. The quantity $f_e$ shows why splitting by color may be necessary. Since $\gammaIAb$ in the two different source samples should be the same, the excess galaxies in the two samples should have the same properties. The excess object populations in samples {\bf a} and {\bf b} are quite different and thus will have different IA signal unless all galaxies, regardless of color, have the same IA behavior.

The behavior of $f_t$ and $f_e$ seen in figure~\ref{fig:eff_frac} is easily understood in terms of basic source clustering properties and photo-$z$ uncertainties. As with many photo-$z$ estimation procedures, ZEBRA provides more accurate redshifts for red galaxies. Poorer photo-$z$ precision for blue objects results in a larger fraction of excess galaxies (actually located at $\zl$) being scattered into the {\bf b} sample. This effect increases the values of $f_r$ and $f_t$ for the {\bf a} sample, since blue galaxies have been preferentially removed. Furthermore, the fact that red galaxies are more highly biased than blue galaxies yields the observed scale dependence: $f_r$ and $f_t$ increase on small scales where clustering is more significant. The effects of this photo-$z$ trend can also be seen in figure~\ref{fig:boosts}. The separation between $B_a$ and $B_b$ is larger for red galaxies, indicating a more accurate physical distinction between the two sub-samples. Similarly, $B_b$ for blue galaxies is larger than for red galaxies, despite having weaker clustering, because there are more physically associated galaxies with large photo-$z$ error.

To examine the effectiveness of the split, we perform an additional test using the ratio of boost factors between the two samples: $(B_a(r_p)-1)/(B_b(r_p)-1)$. The quantity $(B-1)$ measures the fractional contamination from excess sources and is given by a projection of the lens-source cross-correlation (eq.~\ref{eq:B-1 theory}). Comparison of the observed boost with this prediction could indicate that the ``excess'' sources in samples {\bf a} and {\bf b} are statistically equivalent. In principle, knowledge of the source photometric and spectroscopic redshift distributions (e.g. from a calibration sample) allows modeling of the projection weighting ($\tilde{P}(z_s,z_l)$ in eq.~\ref{eq:B-1 theory}), from which it is possible to de-project the measured boost into a cross-correlation function. However, it is impractical to obtain $\tilde{P}(z_s,z_l)$ with adequate resolution to probe the correlation function on relevant scales ($r_p\lesssim 10$ \hMpc). Furthermore, it is challenging to construct a set of galaxies from the calibration set that reliably measures the photo-$z$ behavior of sources near lens galaxies. Another way to probe the sufficiency of the object split is to look at the ratio of measured $B-1$ values for different samples. Because of large photo-$z$ uncertainties, $\tilde{P}(z_s,z_l)$ will be broad compared with $\xi_{ls}$ and is thus effectively constant in the numerator of eq.~\ref{eq:B-1 theory2}. If the $r_p$ and $z_l$ dependencies of $\xi_{ls}$ separate, we find:
\begin{align}
R(r_p) \equiv \frac{B_a(r_p)-1}{B_a(r_p)-1} \propto \frac{w_{ls_a}(r_p)}{w_{ls_b}(r_p)},
\end{align}
where $w_{ls}(r_p) \equiv \int \rmd\Pi p_s(\Pi)\xi_{ls}(r_p,\Pi)$ is the projected cross-correlation function, and $s_a$ and $s_b$ denote the sources found in samples {\bf a} and {\bf b}. The constant of proportionality involves the ratio of integrals over lens and source distributions as well as the photo-$z$ projection weighting $\tilde{P}$.

If samples {\bf a} and {\bf b} contain a statistically similar collection of sources, this ratio will be roughly constant across the range of scales we wish to probe. Scale-dependence in the ratio could indicate that the samples have a different composition and thus have different scale-dependent biases on the cross correlation.

\begin{figure}[t!]
\begin{center}
\includegraphics[width=\textwidth]{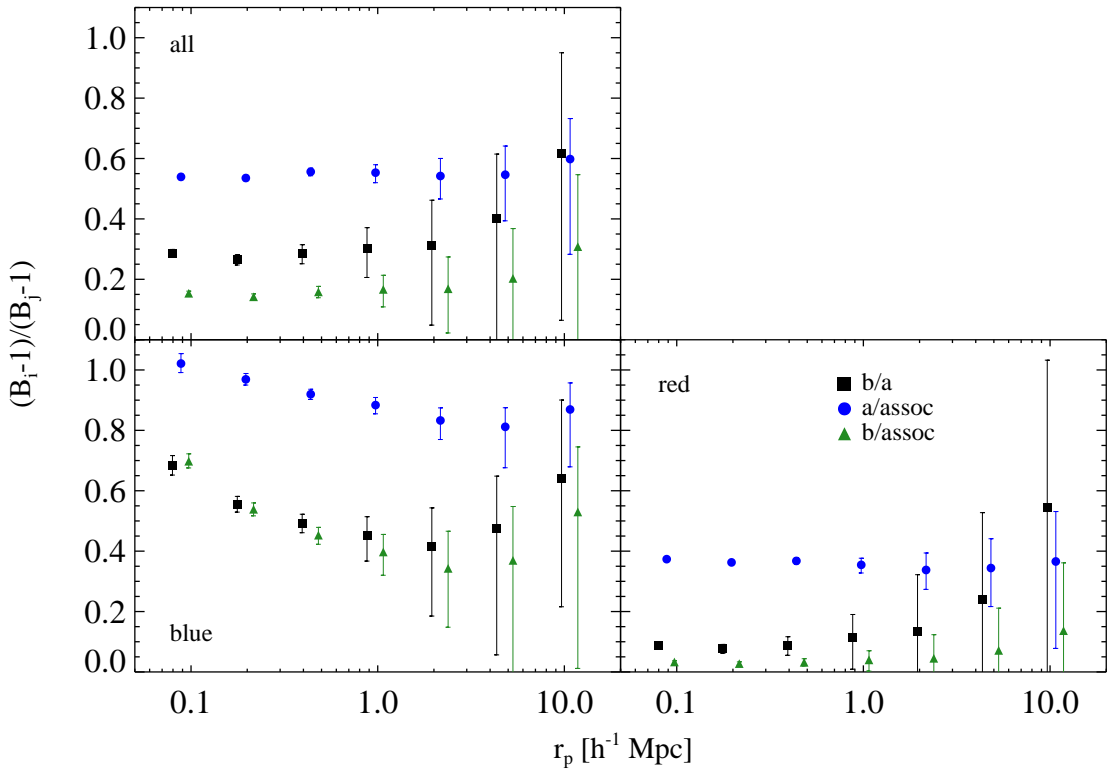}
\end{center}
\caption{$R(r_p)$ is shown for different object colors (as labeled). The ratio is constant, within the estimated uncertainty, for all and red sources but not for blue sources.}
\label{fig:B-1ratio}
\end{figure}

In figure~\ref{fig:B-1ratio}, the quantity $R(r_p)$ is plotted, for all, red, and blue source galaxies.  Previous work indicates that red galaxies have reasonably well-understood patterns of small-scale clustering, photo-$z$ error, and IA \cite{zehavi11,abdalla11,joachimi11}. The roughly constant value of $R(r_p)$ measured for red galaxies supports the fact that we have selected a source sample with uniform properties across the photo-$z$ split. The IA properties of ``blue'' galaxies are poorly understand, allowing for more significant variation within the broad sample. The measurement of $R(r_p)$ for blue sources in figure~\ref{fig:B-1ratio} indicates that the sample exhibits significant heterogeneity in clustering (and likely IA) properties. Results for blue sources are thus more subject to systematic uncertainties, although as discussed above, these uncertainties are currently sub-dominant to the relatively large statistical uncertainty on intrinsic alignment contamination.

However the $R(r_p)$ diagnostic is not conclusive. For instance, the entire source sample with no color split contains multiple galaxy populations, each with different clustering and IA properties. The value of $R(r_p)$ depends on the relative abundance as well as the scale-dependent bias of each. As seen in figure~\ref{fig:B-1ratio}, these factors result in a roughly constant value of $R(r_p)$ for the entire source sample, despite its composite nature. Similarly, if assumptions on the relevant scale dependencies in the measured boost are violated, a uniform sample could exhibit a non-constant $R(r_p)$. We reiterate that for this work, even if the assumption of sample homogeneity is violated, as we expect it to be with the all and blue samples, our method still provides a more self-consistent and unbiased IA measurement than earlier techniques. The bias potentially introduced by the assumption is shown in section~\ref{sec:mimeasure} to be small. In future studies with higher measurement precision, the implications of this assumption should be considered.

More generally, with sufficient statistical power, it will be possible to determine the effect of correlations between IA properties and photometric redshift uncertainties. We expect such correlations to exist, even within a single morphological category, since both IA and photo-$z$ quality are affected by galaxy luminosity (e.g. \cite{hirata07,{feldmann06}}). Splitting the source galaxies into more than two sub-samples will allow a measurement of how $\gammaIAb$ varies. Any two sub-samples can be used to solve for $\gammaIAb$, and the results from different sub-samples can then be compared.

\section{Sources of uncertainty}
\label{sec:uncertainties}
The sources of uncertainty in the IA measurement are seen in eq.~\ref{eq:IAsolution}. $\widetilde{\DS}$ is subject to shape and measurement noise. Scatter in the number of random-source pairs leads to uncertainty in both the boost factors and $\langle \tsigmac \rangle_{\rm ex}$. All of these quantities are affected by photo-$z$ scatter, the bias from which is corrected with the $c_z$ factor. The uncertainty in $c_z$, as measured from calibration sets, is at the $\sim2\text{--}3\%$ level \cite{nakajima11}, including both statistical and systematic effects, and we thus do not include it. We note, however, that $\gammaIAb$ depends only on the ratio of $c_z$ between the source samples. Thus, for deeper surveys for which it can be difficult to obtain sufficiently representative photo-$z$ calibration samples, the uncertainty in $c_z$ can be mitigated by measuring the ratio of the lensing signal for the two samples on large scales ($\approx 50$\hMpc), where the only difference should be due to photo-$z$ calibration. With the exception of $c_z$, all the sources of uncertainty mentioned here are automatically included in the bootstrap realizations. On the range of scales considered here, the most significant source of error is the shape and measurement noise in $\widetilde{\DS}$.

As seen in figure~\ref{fig:DS}, $\widetilde{\DS}$ is measured at similar levels of precision for red and blue sources, since they have roughly the same number of lens-source pairs and measurement noise per pair. However, the uncertainty in the IA signal for blue galaxies is higher than for red galaxies. This difference is due to stronger clustering and more precise ZEBRA photo-$z$ measurement for red source galaxies. The denominator of eq.~\ref{eq:IAsolution} converts the difference between $\widetilde{\DS}$ measurements to the IA signal. When $B_a \gg B_b$, the denominator scales as $B_a-1$, and uncertainty is reduced when strong clustering yields a large value of $B_a$. If the difference between $B_a$ and $B_b$ is smaller, as would be the case for large photo-$z$ uncertainties, the denominator is further reduced. These two factors combine to degrade the precision of the IA measurement for blue galaxies.

Examining these sources of uncertainty also allows us to understand the effects of different source splitting on the measurement precision. If boost factors were held fixed and were perfectly known, the error in $\widetilde{\DS}_a-\widetilde{\DS}_b$ would be minimized when the {\bf a} and {\bf b} samples had equal lens-source pair numbers, weighted by $(c_z B\langle\sigmac\rangle)^{-1}$ for each sample. However, changing the location of the photo-$z$ split will affect both the number of lens-source pairs and the overall boost in each sub-sample. Increasing the value of $B_a$ will decrease the overall error for a given measurement uncertainty. Moreover, uncertainty in the boost factors themselves contribute a non-negligible error, which decreases with larger $B_a$ in the relevant regime. Finally, having $(B_a-1) \gg (B_b-1)$ both decreases the uncertainty and avoids a singular IA estimator, which would yield a large, non-Gaussian variance. Our choice of $\Delta z = 0.17$ was motivated by these considerations, although it is not rigorously shown to be the optimal value. We tried other, similar splitting schemes, and found our choice yields smaller uncertainties in $\gammaIAb$.


\pagebreak

\providecommand{\href}[2]{#2}\begingroup\raggedright\endgroup

\end{document}